\documentclass[a4paper,11pt]{article}
\pdfoutput=1 

\usepackage{jinstpub} 

\usepackage{tabularx}
\usepackage{rotating}
\usepackage{caption}

\newcolumntype{C}{>{\centering\arraybackslash}X}

\usepackage{makecell}

\usepackage{cellspace}
\addparagraphcolumntypes{X}

\usepackage{ifthen} 
\newboolean{uprightparticles}
\setboolean{uprightparticles}{false} 

\usepackage{xspace} 
\usepackage{upgreek}

\def\MagUp {\mbox{\em Mag\kern -0.05em Up}\xspace}

\ifthenelse{\boolean{uprightparticles}}%
{

 \def\Pmu         {\ensuremath{\upmu}\xspace}

 \def\PDelta      {\ensuremath{\Delta}\xspace}                 
 \def\PXi      {\ensuremath{\Xi}\xspace}                 
 \def\PLambda      {\ensuremath{\Lambda}\xspace}                 
 \def\PSigma      {\ensuremath{\Sigma}\xspace}                 
 \def\POmega      {\ensuremath{\Omega}\xspace}                 
 \def\PUpsilon      {\ensuremath{\Upsilon}\xspace}                 
 
 \def\PB      {\ensuremath{\mathrm{B}}\xspace}                 
                  
 \def\PD      {\ensuremath{\mathrm{D}}\xspace}

 \def\PK      {\ensuremath{\mathrm{K}}\xspace}

 \def\Pe      {\ensuremath{\mathrm{e}}\xspace}

 \def\Pi      {\ensuremath{\mathrm{i}}\xspace}

}
{

 \def\Pmu         {\ensuremath{\mu}\xspace}

 \mathchardef\PDelta="7101
 \mathchardef\PXi="7104
 \mathchardef\PLambda="7103
 \mathchardef\PSigma="7106
 \mathchardef\POmega="710A
 \mathchardef\PUpsilon="7107
                  
 \def\PB      {\ensuremath{B}\xspace}                 
                  
 \def\PD      {\ensuremath{D}\xspace}

 \def\PK      {\ensuremath{K}\xspace}

 \def\Pe      {\ensuremath{e}\xspace}

 \def\Pi      {\ensuremath{i}\xspace}

}

\makeatletter
\ifcase \@ptsize \relax
  \newcommand{\miniscule}{\@setfontsize\miniscule{4}{5}}
\or
  \newcommand{\miniscule}{\@setfontsize\miniscule{5}{6}}
\or
  \newcommand{\miniscule}{\@setfontsize\miniscule{5}{6}}
\fi
\makeatother

\DeclareRobustCommand{\optbar}[1]{\shortstack{{\miniscule (\rule[.5ex]{1.25em}{.18mm})}
  \\ [-.7ex] $#1$}}

\def\en         {{\ensuremath{\Pe^-}}\xspace}   
\def\ep         {{\ensuremath{\Pe^+}}\xspace}

\def\mup        {{\ensuremath{\Pmu^+}}\xspace}

  \def\Kbar    {{\kern 0.2em\overline{\kern -0.2em \PK}{}}\xspace}

\def\KorKbar    {\kern 0.18em\optbar{\kern -0.18em K}{}\xspace}

  \def\Dbar    {{\kern 0.2em\overline{\kern -0.2em \PD}{}}\xspace}

\def\DorDbar    {\kern 0.18em\optbar{\kern -0.18em D}{}\xspace}

\def\Bbar    {{\ensuremath{\kern 0.18em\overline{\kern -0.18em \PB}{}}}\xspace}

\def\BorBbar    {\kern 0.18em\optbar{\kern -0.18em B}{}\xspace}

  \def\Y#1S{\ensuremath{\PUpsilon{(#1S)}}\xspace}

\def\Lbar        {{\ensuremath{\kern 0.1em\overline{\kern -0.1em\PLambda}}}\xspace}
\def\LorLbar    {\kern 0.18em\optbar{\kern -0.18em \PLambda}{}\xspace}

\def\order   {{\ensuremath{\mathcal{O}}}\xspace}

\def\AT#1     {\ensuremath{A_{\mathrm{T}}^{#1}}\xspace}           

\def\C#1      {\ensuremath{\mathcal{C}_{#1}}\xspace}                       
\def\Cp#1     {\ensuremath{\mathcal{C}_{#1}^{'}}\xspace}                    
\def\Ceff#1   {\ensuremath{\mathcal{C}_{#1}^{\mathrm{(eff)}}}\xspace}        
\def\Cpeff#1  {\ensuremath{\mathcal{C}_{#1}^{'\mathrm{(eff)}}}\xspace}       
\def\Ope#1    {\ensuremath{\mathcal{O}_{#1}}\xspace}                       
\def\Opep#1   {\ensuremath{\mathcal{O}_{#1}^{'}}\xspace}                    

\newcommand{\tev}{\ifthenelse{\boolean{inbibliography}}{\ensuremath{~T\kern -0.05em eV}}{\ensuremath{\mathrm{\,Te\kern -0.1em V}}}\xspace}
\newcommand{\gev}{\ensuremath{\mathrm{\,Ge\kern -0.1em V}}\xspace}
\newcommand{\mev}{\ensuremath{\mathrm{\,Me\kern -0.1em V}}\xspace}
\newcommand{\kev}{\ensuremath{\mathrm{\,ke\kern -0.1em V}}\xspace}
\newcommand{\ev}{\ensuremath{\mathrm{\,e\kern -0.1em V}}\xspace}
\newcommand{\gevc}{\ensuremath{{\mathrm{\,Ge\kern -0.1em V\!/}c}}\xspace}
\newcommand{\mevc}{\ensuremath{{\mathrm{\,Me\kern -0.1em V\!/}c}}\xspace}
\newcommand{\gevcc}{\ensuremath{{\mathrm{\,Ge\kern -0.1em V\!/}c^2}}\xspace}
\newcommand{\gevgevcccc}{\ensuremath{{\mathrm{\,Ge\kern -0.1em V^2\!/}c^4}}\xspace}
\newcommand{\mevcc}{\ensuremath{{\mathrm{\,Me\kern -0.1em V\!/}c^2}}\xspace}

\def\mus  {\ensuremath{{\,\upmu{\mathrm{ s}}}}\xspace}

\def\order{{\ensuremath{\mathcal{O}}}\xspace}

\def\gsim{{~\raise.15em\hbox{$>$}\kern-.85em
          \lower.35em\hbox{$\sim$}~}\xspace}
\def\lsim{{~\raise.15em\hbox{$<$}\kern-.85em
          \lower.35em\hbox{$\sim$}~}\xspace}

\def\tell1  {TELL1\xspace}
\def\ukl1   {UKL1\xspace}

\title{Real-time data processing with GPUs in high energy physics}

\author{D. vom Bruch}
\affiliation{LPNHE, Sorbonne Universit\'{e}, Paris Diderot Sorbonne Paris Cit\'{e}, CNRS/IN2P3,\\
  4 Place Jussieu, 75252 Paris CEDEX 05, France}

\emailAdd{dovombru@cern.ch}

\abstract{As high energy physics experiments reach higher luminosities and intensities, the computing burden 
for real time data processing and reduction grows. Following the developments in the computing 
landscape, multi-core processors such as graphics processing units (GPUs) are increasingly used for such tasks. These proceedings provide an introduction to the GPU architecture and describe how it maps to common 
tasks in real time data processing. In addition, specific use cases of GPUs in the trigger systems of five different high energy physics
experiments are presented.}


\arxivnumber{} 

\proceeding{INFIERI2019}

\begin{document}
\maketitle
\flushbottom

\section{Computing challenge in high energy physics}
\label{sec:computing_challenge}
In the quest for exploring new physics scenarios in high energy physics (HEP) the collection
of high statistics data sets is crucial. Therefore, experimental facilities are designed for increasing
luminosities and beam intensities. This goes in line with increasing data rates and higher 
computing demands to process the data.
In Run~3 of the LHC, beginning in 2021, LHCb and ALICE will process
 data rates of 40~Tbit/s and 30~Tbit/s respectively in software. At the software level trigger, ATLAS and CMS will reach this order of 
magnitude in Run~4, beginning in 2027. Not only LHC experiments, but also dedicated 
fixed-target experiments face increasingly higher decay and data rates in the future,
 such as the Mu3e experiment at the Paul Scherrer Institute and NA62 at CERN.

\begin{figure}[htbp]
\centering
\includegraphics[width=\textwidth]{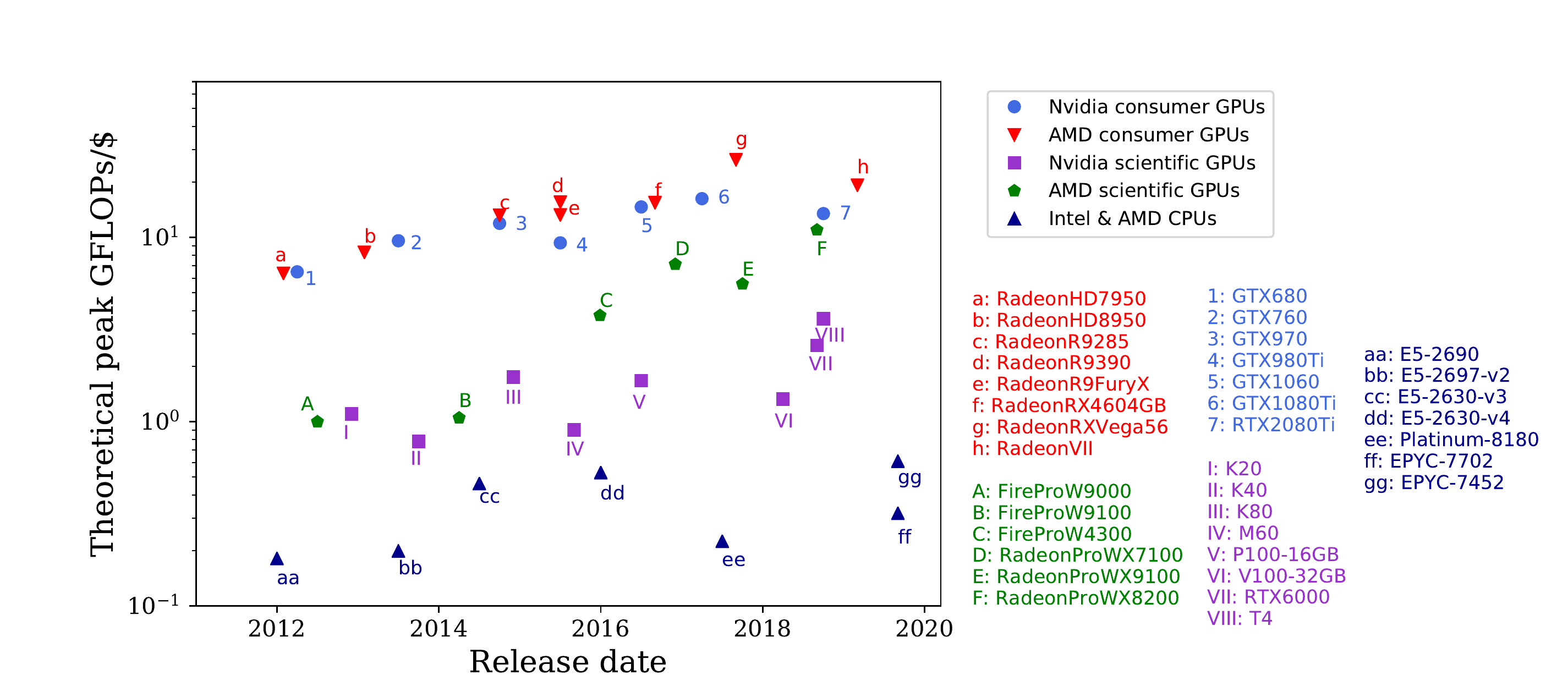}
\caption{\label{fig:teraflops_dollar} Theoretical peak performance per dollar (launch price) versus release date for (red downward triangles) consumer AMD GPUs, (blue circles) consumer Nvidia GPUs, (green pentagons) scientific AMD GPUs, (purple squares) scientific Nvidia GPUs and (dark blue upward triangles) Intel and AMD CPUs.}
\end{figure}

As the computing landscape is changing, the hardware used for real-time data selection needs to be adopted. 
With the stagnation of the single thread computing performance since the mid-2000s, the 
growing computing demand cannot be met with traditional single-core CPU processors. 
What is more, the prediction typically assumed in high energy physics (HEP) that the same amount of money will 
buy hardware with 15-20\% better performance in the next year (``flat budget'') does not 
hold without adopting new emerging architectures and computing models. Therefore, the importance 
of multi-core architectures has increased significantly over the last years. 
One particular parallel processor provides thousands of cores and tens of TFLOPs in 
processing power for highly parallelizable tasks: the graphics processing unit (GPU).
When fully utilized, a GPU offers the best TFLOPS/\$ performance, as illustrated in Figure~\ref{fig:teraflops_dollar}. The price-performance advantage is especially pronounced for consumer GPU cards, but also scientific cards offer a higher performance per dollar as compared to CPUs. Furthermore, the price-performance increases more over time for GPUs than for CPUs.

 In HEP, the task of real-time data processing and reduction (``trigger'') poses 
a significant compute load, since particle decays in the detector need to be 
reconstructed partially or fully to efficiently select signal decays of interest. 
In these proceedings, I discuss the usage of GPUs in the context of real-time data 
processing in HEP. In section~\ref{sec:gpu_computing}, the GPU architecture is introduced 
and compared to other processors,
 section~\ref{sec:trigger_tasks} covers the typical tasks and algorithms in trigger applications and how they fit the GPU architecture. Finally, in section~\ref{sec:GPUs_experiments} examples are discussed of how GPUs are used in data processing at different experiments.

\section{Introduction to GPU computing}
\label{sec:gpu_computing}

GPUs are designed to display graphics on a computer screen. To transform the color and shade of every object onto the millions of pixels of a screen, the hardware prioritizes throughput over latency.
Since the mid-2000s, programmable GPU processors 
can be used for general purpose GPU computing.

\subsection{The GPU architecture}
\begin{figure}[htbp]
\centering 
\includegraphics[width=.7\textwidth]{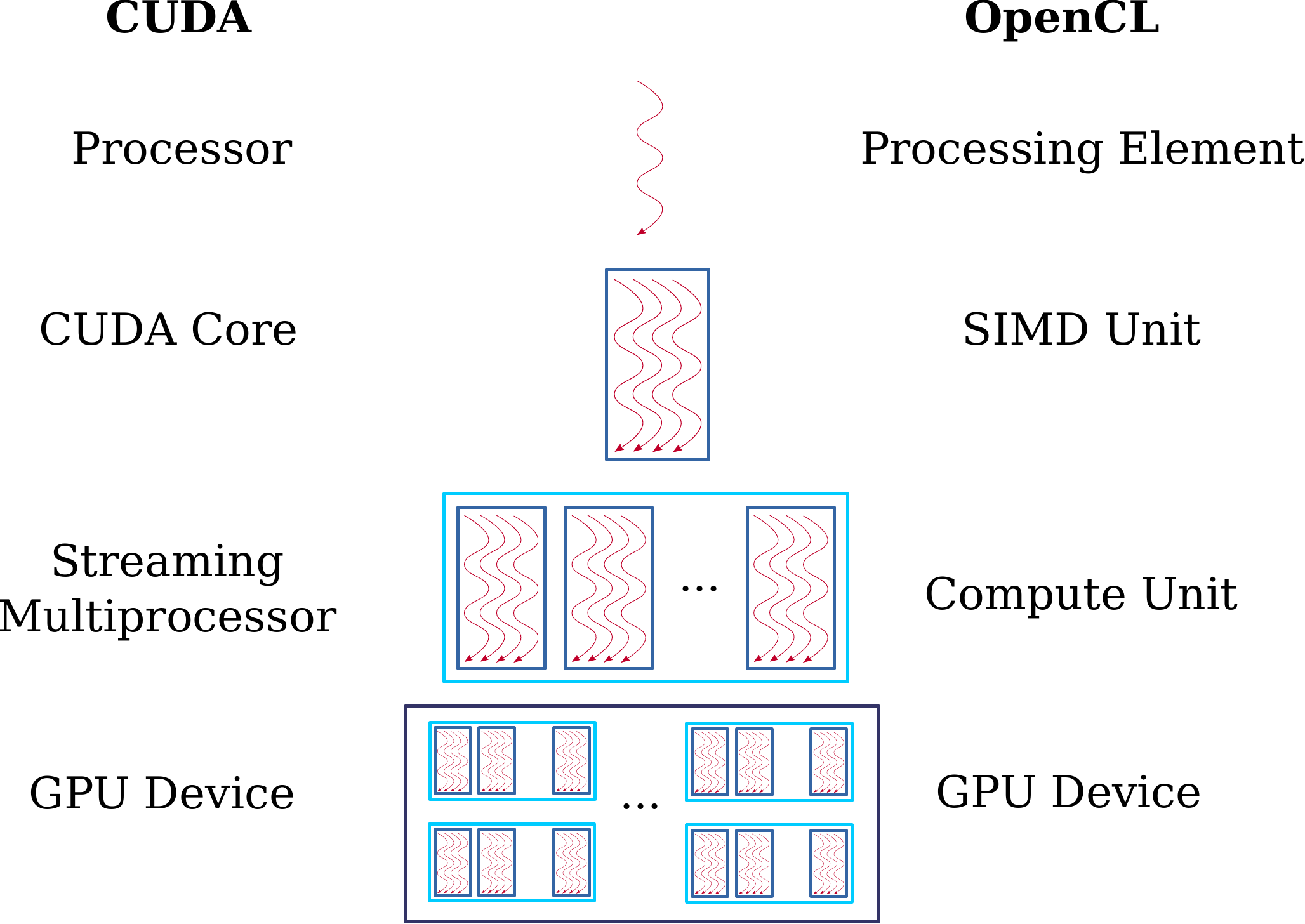}
\caption{\label{fig:gpu_architecture} Terminology to describe the GPU architecture with CUDA and OpenCL.}
\end{figure}

The GPU computing paradigm follows 
the Single Instruction Multiple Threads (SIMT) approach, which has similarities with the Single Instruction Multiple Data (SIMD) paradigm used for vector processing on CPUs. In SIMT, a single instruction decoder is available for multiple threads, processing in lock-step. 
One algorithm, the so called ``kernel'', is executed on many threads, and every thread processes independent data sets. Groups of threads make up a core and the cores are again grouped 
into units that then make up the GPU.
The two most commonly used frameworks for programming GPUs are CUDA~\cite{cuda}, developed by Nvidia for their GPUs, 
and the cross-platform standard OpenCL~\cite{opencl} maintained by the Khronos Group, supporting both AMD and Nvidia GPUs, as well as other hardware accelerators. In addition, the open source project HIP~\cite{hip} is being developed to support both AMD and Nvidia GPUs and several cross-architecture frameworks such as SYCL~\cite{sycl} (also maintained by the Khronos Group and building on the concepts of OpenCL) and Alpaka~\cite{alpaka} are emerging to allow software development for various back ends.
Both CUDA and OpenCL have distinct terminology 
to describe the GPU architecture, as illustrated in Figure~\ref{fig:gpu_architecture}. 
Threads in a Streaming Multiprocessor or Compute Unit share a set of resources such 
as single and double precision cores, function units and memory.

In the software abstraction for the hardware illustrated in Figure~\ref{fig:gpu_architecture},
 the parallelization of an algorithm is assigned on two levels: 
threads are grouped into blocks and many blocks of threads make up a
grid. Threads within one block share a common memory and can be synchronized. 
All threads in one block are always assigned to the same Streaming Multiprocessor / 
Compute Unit. Given the hardware constraints, one has to optimize the resources needed by 
a kernel, as well as the number of threads per block and blocks per grid to achieve an 
efficient usage of the GPU processing power.

\subsection{I/O of a GPU}
A GPU is typically connected to the host CPU via a PCIe connection. 
All data processed on the GPU is copied from the CPU to GPU memory and any
 results required on the CPU are copied back via this PCIe connection. It is therefore 
crucial to verify that the PCIe bandwidth does not pose a limitation to using the GPU as 
accelerator. Current GPU models are equipped with PCIe~3.0 connections (16 GB/s for 16 lanes), 
while the next generation of cards is foreseen to support PCIe~4.0 (32 GB/s for 16 lanes). Scientific Nvidia GPUs also 
provide the Nvlink protocol with a maximum data rate of up to 100 GB/s as interconnect among GPUs.

\subsection{GPUs compared to other processors}
As GPUs are designed to process the same arithmetic on independent data, they are 
optimal at parallel performance. Compared to CPUs, the GPU cache is smaller with higher 
latency, the processor runs at lower frequency and there are no speculative executions. 
However, by scheduling the thousands of threads optimally, the GPU cores always have 
work to do and hide the latency via high throughput. 
Apart from CPUs and GPUs, field programmable gate arrays (FPGAs) are typically used in the data acquisition of HEP experiments.
 With their fixed, short latency and versatile I/O connectors they are well suited for the
 early stages of the readout chain. A summary of the different characteristics of CPUs, GPUs 
and FPGAs is listed in table~\ref{tab:compare_processors}.

\begin{table}[!htbp]
\centering%
\caption{\label{tab:compare_processors} GPU characteristics as compared to CPUs and FPGAs.}
\smallskip
\begin{tabularx}{\linewidth}{|C|C|C|C|}
\hline
& \textbf{CPU} & \textbf{GPU} & \textbf{FPGA}\\
\hline
\hline
\textbf{Latency} & \order(10)~$\mus$ & \order(100)~$\mus$ & Deterministic, \order~(100)~ns \\
\hline
\textbf{I/O with processor} & Ethernet, USB, PCIe & PCIe, Nvlink & Connectivity to any data source via printed circuit board (PCB)\\
\hline
\textbf{Engineering cost} & Low entry level (programmable with c++, python, etc.) & Low entry level (programmable with CUDA, OpenCL, etc.) & Some high-level syntax available, traditionally VHDL, Verilog (specialized engineer) \\
\hline
\textbf{Single precision floating point performance} & \order(10)~TFLOPs & \order(10)~TFLOPs & Optimized for fixed point performance \\
\hline
\textbf{Serial / parallel} & Optimized for serial performance, increasingly using vector processing  & Optimized for parallel performance & Optimized for parallel performance \\
\hline
\textbf{Memory} & \order(100)~GB RAM & \order(10)~GB & \order(10)~MB (on the FPGA itself, not the PCB) \\
\hline
\textbf{Backward compatibility} & Compatible, except for vector instruction sets & Compatible, except for specific features only available on modern GPUs & Not easily backward compatible\\
\hline
\end{tabularx}
\end{table}

\section{Real-time data processing tasks}
\label{sec:trigger_tasks}
Over the past decade, GPUs have found more and more use cases within the field of HEP. 
They are used in data analysis and simulation, mainly through machine learning tools. 
In these proceedings, the focus lies on the usage of GPUs for data reduction via real-time analysis.

\subsection{Data reduction strategies}
Data reduction in HEP can be divided into two main categories: a selection is either
 possible based on information from specific detectors or detector regions (``local''), or 
the information from several detectors is combined (``global''). If the decays of interest generate local characteristic 
signatures, such as energy deposit in the calorimeter, the former method can be used. 
The general purpose detectors ATLAS and CMS fall into this category with their main interest
 lying in Higgs, jet and electroweak physics. Local selections are also necessary if the data 
stream is too large to be read out from the detector completely. In this case, low latency 
and high bandwidth are required, such that FPGAs or custom circuit boards (ASICS) are best
 suited and the type of 
selection is referred to as ``hardware level trigger''\footnote{In some cases, the information from several sub-detectors is combined even at the hardware level trigger, these are then also referred to as ``global``.}.
Global selections on the other hand are possible if the whole data stream can be read 
out or has already been reduced by a hardware level trigger. They are especially useful if signal 
decays are in large abundance or highly resemble background processes. In this case, the 
characteristics of the decays are determined by reconstructing the particle 
trajectories within the detector (``track reconstruction'') and possibly performing particle identification and / or adding information from the calorimeters. Since the full data stream has 
already been read out at this stage, the latency requirements are relaxed compared to 
the hardware level trigger. Consequently, this type of selection is referred to as  ``software level trigger'' and processors such as CPUs and GPUs can be used. 

\subsection{Mapping real time analysis tasks to GPUs}
To determine whether the compute performance of a GPU can efficiently be exploited for real time analysis, it is crucial to establish how many tasks are ``parallelizable``. A ``parallelizable`` task is one where the same arithmetic acts on independent sets of data. Only the parallel part of a program benefits from the speedup due to many processors, as stated in 
 Amdahl's law~\cite{amdahl}. Therefore, only problems and algorithms with large portions of parallel tasks 
can make use of the processing power of a GPU.

Real-time data reduction in software typically contains several or all of the following tasks:

\begin{itemize}
\item Decoding the raw input into the global coordinate system of an experiment;
\item Clustering of measurements caused by the passage of the same particle in 
one detector unit into single coordinates (``hits'');
\item Finding combinations of hits originating from the same particle trajectory (pattern
 recognition);
\item Describing the track candidates from the pattern recognition step with a track model (track fitting);
\item Reconstructing primary and secondary vertices from the fitted tracks (vertex finding);
\item Performing particle identification with dedicated detectors;
\item Reconstructing the shower caused by a particle in the calorimeter;
\item Applying selections to the reconstructed candidates.
\end{itemize}

In addition to processing many particle collisions (or time slices in the case of continuous 
beam experiments) in
 parallel, the above tasks are also parallelizable. The decoding of raw input factorizes by
 readout unit. Clustering in tracking detectors or calorimeters can be performed in contained regions of a detector. 
The main compute burden of pattern recognition is the combinatorics of the many possible
 hit combinations, which can also be processed in parallel. Finally, tracks can be fitted
 independently from one another, as well as the different combinations of tracks forming a 
vertex. Particle identification can typically be processed per candidate track. It is often convenient to map the execution of a specific algorithm for one event or time slice to a block of threads, as communication among threads is possible in this case. In the case of large events, the processing of data in a sub-detector may be split into several blocks based on the geometry of the detector.

The various possibilities for parallelization within the software level trigger make all 
of its tasks or at least a few compute intensive ones optimal candidates to be processed on 
GPUs.

\section{GPU usage in real-time analysis at HEP experiments}
\label{sec:GPUs_experiments}

Various experiments in HEP consider the usage of GPUs for real-time data reduction or 
have already employed GPUs in the trigger in the past. 
Especially track reconstruction is highly compute intensive, so this task is a typical candidate to be processed on GPUs. 
The following sub-sections describe five different approaches of using GPUs at the trigger level, first for the fixed target 
experiments NA62 and Mu3e and then for the three LHC experiments CMS, ALICE and LHCb.
A common feature among all described use cases is the coherence of the work flow on the GPU itself, as it reduces the amount of memory copies required between the GPU and the host CPU.
 A comparison of the GPU usage in the different experiments is summarized in
 table~\ref{tab:experiments_gpus}.

\begin{sidewaystable}[!htbp]
\centering%
\caption{\label{tab:experiments_gpus} Overview of GPU usage for real-time analysis in various HEP experiments.}
\smallskip
\begin{tabularx}{\linewidth}{|C|C|C|C|C|C|C|}
\hline
\textbf{Experiment} &  \textbf{Main task processed on GPU} & \textbf{Event / data rate} & \textbf{Number of GPUs} & \textbf{Types of GPUs tested} & \textbf{Date for employment} & \textbf{References}\\
\hline
\hline
\textbf{NA62} &  RICH ring pattern reconstruction & 10~MHz / 2.5~Gbit/s & 1 & Nvidia K20c, P100 & Tested in 2017 \& 2018, planned for 2021 & \cite{Ammendola2016,Ammendola2018}\\
\hline
\textbf{Mu3e} &  Track- \& vertex reconstruction in the pixel tracker, data selection & 20~MHz / 32~Gbit/s & \order(10) & Nvidia GTX980, GTX1080, RTX1080Ti & 2021 & \cite{VomBruch2017}\\
\hline
\textbf{CMS} &  Decoding of raw data, clustering, pattern recognition in the pixel detector & 100~kHz / -  & & Nvidia RTX2080, K20 &  Planned for 2021 & \cite{Funke2014, Pantaleo2016}\\
\hline
\textbf{ALICE} &  Track reconstruction in the TPC & <~500~Hz Pb-Pb or <~2~kHz p-p / <~100~Gbit/s & 64 & Nvidia GTX480 & 2010-2013 & \cite{fe0cd925bfcd4edd92b06d6cb279d97a} \\
\hline
\textbf{ALICE} &  Track reconstruction in the TPC & <~1~kHz Pb-Pb or <~2~kHz p-p / <~384~Gbit/s & 180 & AMD S9000 & 2015-2018 & \cite{fe0cd925bfcd4edd92b06d6cb279d97a} \\
\hline
\textbf{ALICE} &  Track reconstruction in three sub-detectors & 50~kHz Pb-Pb or <5~MHz p-p / 30~Tbit/s & \order(2000) & & 2021 & \cite{refId0,Buncic:2011297}\\
\hline
\textbf{LHCb} &  Decoding of raw data, clustering, track reconstruction in three sub-detectors, vertex reconstruction, muon identification, inclusive selections & 30~MHz / 40~Tbit/s &  \order(500) & Nvidia RTX2080Ti, RTX6000, V100& Possibly in 2021 & \cite{allen_publication} \\
\hline
\end{tabularx}
\end{sidewaystable}

\subsection{NA62}
The NA62 experiment at CERN is dedicated to the study of rare kaon decays. 
During the low level trigger, the event rate is reduced from 10~MHz to 1~MHz and
muon-pion particle identification occurs via a Ring Imaging Cherenkov Detector (RICH).
 An R\&D project exists to perform the reconstruction of the ring-shaped patterns in the 
RICH detectors on GPUs already at the first trigger level~\cite{Ammendola2016}.
 Rings are reconstructed by either filling histograms of distances from every photo 
multiplier to measurements in parallel and finding the one that best fits a circle or by 
making use of the Almagest algorithm~\cite{Lamanna2014}. The challenge in using a GPU at the earliest 
trigger stage lies in meeting the strict latency requirements. Therefore, a dedicated network 
interface card was designed to handle the data transfers to and from the GPU directly 
over the PCIe switch~\cite{Ammendola2018}. With this setup, receiving the data, sending it to the GPU from the network interface card, processing on the GPU and the transfer back take at most 350~$\mus$~\cite{Ammendola2016}. A test bed processing data at 5-6~MHz was installed during 2017 and 2018 data taking, the GPU reconstruction is planned to run at 10~MHz in 2021.

\subsection{Mu3e}
Designed for the search of the lepton flavour violating decay
 $\mup \rightarrow \ep\en\ep$, the Mu3e experiment is being constructed at the 
Paul Scherrer Institute in Switzerland. The complete data-stream of 80~Gbit/s is read out and split into 50~ns time slices. The data selection will occur fully on GPUs 
based solely on data from the central pixel detector.  Combinations of three hits are already determined in the readout FPGA board and transferred to the GPU at a data rate of 32~Gbit/s.
Then the 3-hit stubs are extended to the fourth pixel layer and the linear three-dimensional track 
fit with multiple scattering developed for Mu3e~\cite{Berger2017} is processed. 
Finally, three track vertices are reconstructed based on geometric constraints and selection
decisions are copied back from the GPU to the host CPU. Processing time slices and track seeds in parallel,
12 GTX 1080 Ti GPUs cards are sufficient to process the full data stream and reduce the 
event rate by a factor 100~\cite{VomBruch2017}. 

\subsection{CMS}
To cope with a pile-up of 140 at the high luminosity LHC (Run~4, starting in 2027), CMS plans to 
introduce a new trigger stage for track reconstruction using high performance computing platforms.
In this context track reconstruction of the pixel detector
is proposed to run on GPUs~\cite{Funke2014, Pantaleo2016} within the high level trigger. At this point, the event rate has already been reduced to 100~kHz by the hardware level trigger. Decoding of the pixel raw data, clustering and pattern recognition are off-loaded per event to a GPU coprocessor. The cellular automaton algorithm~\cite{etde_20318648} is used for pattern recognition. Based on a graph of interconnected cells, this algorithm is easily parallelizable as the cells (segments of a track) can be built independently. Offloading the pixel track reconstruction to GPUs in the high level trigger is already planned for Run~3 of the LHC in 2021.

\subsection{ALICE}
At ALICE, GPUs have been used in the high level trigger~\cite{fe0cd925bfcd4edd92b06d6cb279d97a} to perform track reconstruction within the
 time projection chamber (TPC) for calibration purposes already since Run~1 of the 
LHC~\cite{Rohr_2012}. Since the primary goal of ALICE is studying the quark-gluon plasma in Pb-Pb collisions,
 the event size is orders of magnitude larger than for other LHC experiments, but events occur at a lower rate. As opposed to the other experiments described in these proceedings, ALICE compresses its data rather than selecting it. For the compression, high level objects such as reconstructed tracks are required. Therefore, the steps are similar to those used in other experiments for selection.
 Similarly to the CMS approach, 
the cellular automaton algorithm is used for pattern recognition to find track seeds within the TPC. In addition, a Kalman filter~\cite{Kalman1960} is employed for track forwarding and the track fit. 
During Runs~1 and~2 of the LHC, a hardware level trigger was used, followed by calibration and compression in the high level trigger.
In Run~3, ALICE switches to a fully software trigger scheme and the data rate processed in the high level trigger
will increase from 384~Gbit/s in Run~2 to 30~Tbit/s. The GPU algorithms are updated for the higher event rate and for the upgraded TPC detector. In addition, other parts of the event selection might be processed on the 
GPU in addition to the TPC track reconstruction, such as the extension of tracks inside the TPC to the Inner tracking System (ITS), consisting of pixel detectors~\cite{refId0,Buncic:2011297}.

\subsection{LHCb}
The LHCb experiment is designed for the study of beauty and charm quarks. Part of the 
extensive detector upgrade ongoing for Run~3 is the complete readout of all detectors at 
40~Tbit/s and an entirely software based trigger, split into two stages. During the first stage, 
High Level Trigger 1 (HLT1), the event rate is reduced from 30~MHz by a factor 30-60 based 
on inclusive 1- and 2-track selections. The second trigger 
stage on the other hand is mainly based on exclusive selections. The baseline 
design of the data acquisition system foresees the two trigger stages to be processed 
entirely by CPUs~\cite{CERN-LHCC-2014-016}. An alternative approach has also been developed in the so called 
``Allen'' project, where the full HLT1 sequence was implemented to run on 
GPUs~\cite{allen_publication}. The concept is based on transferring raw data to the GPU, 
processing everything from decoding the binary payload to event selections on the GPU, 
and copying only the decisions and underlying objects back to the host CPU. 
Decoding for four sub-detectors, clustering in the pixel detector, track reconstruction in 
three sub-detectors, muon identification, as well as primary- and secondary vertex reconstruction and the selections are implemented on the GPU.
 The full data stream is 
processed on fewer than 500 Nvidia V100, Quadro RTX 6000 or RTX 2080 Ti cards. 

In the baseline solution, two distinct server farms handle the data stream: one that receives data from the different sub-detectors and builds events and a second one where both high level trigger stages are executed. The data stream is only reduced in the second server farm in this scenario. Since 500 GPU cards physically fit into the first server farm, the data rate can already be reduced at an earlier level if HLT1 is executed on GPUs. Therefore, a significantly cheaper network is required between the first and second server farm and money can instead be spent on GPUs. This demonstrates that GPUs naturally integrate into LHCb's data acquisition system and make it more compact.

\section{Conclusion}
The computing demand of real-time data processing is increasing with higher
 luminosities and beam intensities in HEP. To address this challenge in view of
 the changing computing landscape, parallel processors such as GPUs are emerging in trigger 
systems due to their high price-performance. The strength of GPUs lies in processing the same computation on independent data. This concept matches well to algorithms used in real-time analysis, such as the reconstruction of
particle trajectories.
GPU cards are either employed
for performing a specific task of data processing or to handle a full trigger stage, mostly at the 
level of software triggers.
Numerous HEP experiments already use or plan to use
GPUs both at colliders and in fixed target facilities.
These developments will likely impact the design of trigger systems at future experiments and facilities.

\acknowledgments
D. vom Bruch acknowledges support of the European Research Council Consolidator 
grant RECEPT 724777. The author thanks the organizers of the
 INFIERI Summer School, especially Aurore Savoy-Navarro and Nicola D'Ascenzo, for 
making this excellent school happen. The author would also like to thank Daniel Campora, Vava Gligorov,  Gianluca Lamanna, Felice Pantaleo and David Rohr for fruitful discussions and careful reading of these proceedings.

\bibliographystyle{JHEP}
\bibliography{main}   

\providecommand{\href}[2]{#2}\begingroup\raggedright\begin{thebibliography}{10}

\bibitem{cuda}
``{CUDA Toolkit}.'' \url{https://docs.nvidia.com/cuda/}.

\bibitem{opencl}
``{OpenCL Framework}.'' \url{https://www.khronos.org/opencl/}.

\bibitem{hip}
``{HIP: C++ Heterogeneous-Compute Interface for Portability}.''
  \url{https://gpuopen.com/compute-product/hip-convert-cuda-to-portable-c-code/}.

\bibitem{sycl}
``{SYCL: C++ Single-source Heterogeneous Programming for OpenCL}.''
  \url{https://www.khronos.org/sycl/}.

\bibitem{alpaka}
E.~{Zenker}, B.~{Worpitz}, R.~{Widera}, A.~{Huebl}, G.~{Juckeland},
  A.~{Knüpfer} et~al., \emph{Alpaka -- an abstraction library for parallel
  kernel acceleration},
  \href{http://dx.doi.org/10.1109/IPDPSW.2016.50}{\emph{2016 IEEE International
  Parallel and Distributed Processing Symposium Workshops (IPDPSW)} (2016)
  631--640}.

\bibitem{amdahl}
G.~M. Amdahl, \emph{Validity of the single processor approach to achieving
  large scale computing capabilities},
  \href{http://dx.doi.org/10.1145/1465482.1465560}{\emph{Proceedings of the
  April 18-20, 1967, Spring Joint Computer Conference} (1967) 483--485}.

\bibitem{Ammendola2016}
R.~Ammendola, A.~Biagioni, P.~Cretaro, S.~{Di Lorenzo}, R.~Fantechi, M.~Fiorini
  et~al., \emph{{GPU-based Real-time Triggering in the NA62 Experiment}},
  2016.
\newblock 10.1088/1748-0221/14/04/P04013.

\bibitem{Ammendola2018}
R.~Ammendola, M.~Barbanera, A.~Biagioni, P.~Cretaro, O.~Frezza, G.~Lamanna
  et~al., \emph{{Real-time heterogeneous stream processing with NaNet in the
  NA62 experiment}},
  \href{http://dx.doi.org/10.1088/1742-6596/1085/3/032022}{\emph{J. Phys}
  (2018) 32022}.

\bibitem{VomBruch2017}
D.~vom Bruch, \emph{{Online Data Reduction using Track and Vertex
  Reconstruction on GPUs for the Mu3e Experiment}},
  \href{http://dx.doi.org/10.1051/epjconf/201715000013}{\emph{EPJ Web of
  Conferences} {\bfseries 150} (2017) 00013}.

\bibitem{Funke2014}
D.~Funke, T.~Hauth, V.~Innocente, G.~Quast, P.~Sanders and D.~Schieferdecker,
  \emph{{Parallel track reconstruction in CMS using the cellular automaton
  approach}},
  \href{http://dx.doi.org/10.1088/1742-6596/513/5/052010}{\emph{Journal of
  Physics: Conference Series} {\bfseries 513} (2014) 052010}.

\bibitem{Pantaleo2016}
F.~Pantaleo, G.~Cappello, B.~Hegner, V.~Innocente, A.~B. Meyer, A.~Pfeiffer
  et~al., \emph{{Development of a phase-II track trigger based on GPUs for the
  CMS experiment}},
  \href{http://dx.doi.org/10.1109/NSSMIC.2015.7581775}{\emph{2015 IEEE Nuclear
  Science Symposium and Medical Imaging Conference, NSS/MIC 2015} (2016) 1--6}.

\bibitem{fe0cd925bfcd4edd92b06d6cb279d97a}
{{ALICE Collaboration}}, \emph{Real-time data processing in the alice high
  level trigger at the lhc},
  \href{http://dx.doi.org/10.1016/j.cpc.2019.04.011}{\emph{Computer Physics
  Communications} {\bfseries 242} (2019) 25--48}.

\bibitem{refId0}
{Rohr, David}, {Gorbunov, Sergey}, {Ole Marten, Schmidt} and {Shahoyan, Ruben},
  \emph{Gpu-based online track reconstruction for the alice tpc in run 3 with
  continuous read-out},
  \href{http://dx.doi.org/10.1051/epjconf/201921401050}{\emph{EPJ Web Conf.}
  {\bfseries 214} (2019) 01050}.

\bibitem{Buncic:2011297}
P.~Buncic, M.~Krzewicki and P.~Vande~Vyvre, \emph{{Technical Design Report for
  the Upgrade of the Online-Offline Computing System}},  Tech. Rep.
  CERN-LHCC-2015-006. ALICE-TDR-019, 2015.

\bibitem{allen_publication}
{R. Aaij, J. Albrecht, M. Belous, P. Billoir, T. Boettcher, A. Brea
  Rodr\'iguez, D. vom Bruch, D. H. C\'ampora Pérez, A. Casais Vidal, D. C.
  Craik, P. Fernandez Declara, L. Funke, V. V. Gligorov, B. Jashal, N. Kazeev,
  D. Mart\'inez Santos, F. Pisani, D. Pliushchenko, S. Popov, R. Quagliani, M.
  Rangel, F. Reiss, C. S\'anchez Mayordomo, R. Schwemmer, M. Sokoloff, H.
  Stevens, A. Ustyuzhanin, X. Vilas\'is Cardona, M. Williams}, \emph{{Allen: A
  high level trigger on GPUs for LHCb}}, {\emph{submitted to Computing and
  Software for Big Science} (2019) },
  [\href{https://arxiv.org/abs/1912.09161}{{\ttfamily 1912.09161}}].

\bibitem{Lamanna2014}
G.~Lamanna, \emph{{Almagest, a new trackless ring finding algorithm}},
  \href{http://dx.doi.org/10.1016/j.nima.2014.05.073}{\emph{Nuclear Instruments
  and Methods in Physics Research Section A: Accelerators, Spectrometers,
  Detectors and Associated Equipment} (2014) 241--244}.

\bibitem{Berger2017}
N.~Berger, A.~Kozlinskiy, M.~Kiehn and A.~Sch{\"{o}}ning, \emph{{A new
  three-dimensional track fit with multiple scattering}},
  \href{http://dx.doi.org/10.1016/j.nima.2016.11.012}{\emph{NIM A} {\bfseries
  844} (2017) 135--140},
  [\href{https://arxiv.org/abs/arXiv:1606.04990}{{\ttfamily
  arXiv:1606.04990}}].

\bibitem{etde_20318648}
I.~Abt, D.~Emeliyanov, I.~Kisel and S.~Masciocchi, \emph{Cats: a cellular
  automaton for tracking in silicon for the hera-b vertex detector},
  \href{http://dx.doi.org/10.1016/S0168-9002(02)00790-8}{\emph{Nuclear
  Instruments and Methods in Physics Research. Section A, Accelerators,
  Spectrometers, Detectors and Associated Equipment} {\bfseries 489} (2002) }.

\bibitem{Rohr_2012}
D.~Rohr, S.~Gorbunov, A.~Szostak, M.~Kretz, T.~Kollegger, T.~Breitner et~al.,
  \emph{{ALICE} {HLT} {TPC} tracking of pb-pb events on {GPUs}},
  \href{http://dx.doi.org/10.1088/1742-6596/396/1/012044}{\emph{Journal of
  Physics: Conference Series} {\bfseries 396} (2012) 012044}.

\bibitem{Kalman1960}
R.~E. Kalman, \emph{A new approach to linear filtering and prediction
  problems}, \href{http://dx.doi.org/0.1115/1.3662552 [doi]}{\emph{Journal of
  Basic Engineering} {\bfseries 82} (1960) 35--45}.

\bibitem{CERN-LHCC-2014-016}
\emph{{LHCb Trigger and Online Upgrade Technical Design Report}},  Tech. Rep.
  CERN-LHCC-2014-016. LHCB-TDR-016, 2014.

\end{thebibliography}\endgroup








\end{document}